\documentclass[aps,prd,groupedaddress,preprint,eqsecnum,nofootinbib]{revtex4}
\usepackage{graphicx,epsf,amssymb,amsbsy,amsfonts,amssymb,amsmath}

\begin{document}
\hfuzz 12pt


\title{Wess-Zumino Consistency Condition for Entanglement Entropy}

\author{Shamik Banerjee}

\affiliation{Stanford Institute for Theoretical Physics and Department of
Physics, Stanford University,
Stanford, CA 94305-4060, USA}
\email{bshamik@stanford.edu}





\begin{abstract}
In this brief note, we consider the variation of the entanglement entropy of a region as the shape of the entangling surface is changed. We show that the variation satisfies a Wess-Zumino like integrability condition in field theories which can be consistently coupled to gravity. In this case the "anomaly" is localized on the entangling surface. The solution of the integrability condition should give all the nontrivial finite local terms which can appear in the variation of the entanglement entropy.
\end{abstract}
\maketitle

\section{Introduction}
Recently much attention has been paid to the computation of Entanglement Entropy in quantum field theories. Entanglement entropy of any quantum system is defined in the following way. Given any system we divide it into two subsystems denoted by A and B. If we assume that the observables which describe subsystem A and subsystem B, commute among themselves then the total Hilbert space of the system can be written as the tensor product of the individual Hilbert spaces of subsystems A and B, $H = H_{A}\otimes H_{B}$. Let us further assume that the total system is described by a density matrix $\rho$. We can define a reduced density matrix for the subsystem A denoted by $\rho_{A}$ as $\rho_{A} = Tr_{H_{B}} \rho$. The entanglement entropy of the subsystem A can be defined as the von Neumann entropy of the reduced density matrix $\rho_{A}$,
\begin{equation}
S_{EE}(A) = -Tr_{H_{A}}\rho_{A} ln\rho_{A}
\end{equation}
Entanglement entropy of the subsystem B can also be defined in the same way. Direct evaluation of the entanglement entropy in field theories is complicated due to the presence of the factor $ln\rho$. One can apply the replica trick to compute the trace but in practice this can be done only for the simplest field theories or theories with very high degree of symmetry. For example, the entanglement entropy is known exactly in $1+1$ dimensional conformal field theories when the subsystems are a line segment and its complement \cite{Calabrese:2004eu, Holzhey:1994we, Calabrese:2009qy}. It can also be calculated exactly in field theories which have a holographic dual description \cite{Ryu:2006bv,Ryu:2006ef, Schwimmer:2008yh,Hung:2011ta, Casini:2011kv, Hung:2011xb, Myers:2010tj, deBoer:2011wk}. For applications to the black hole physics please see \cite{Solodukhin:2006xv,Solodukhin:2008dh,Azeyanagi:2007bj,Sen:2011cn,susskind}.

Although the computation of entanglement entropy is extremely difficult, the final answer is to some extent universal. For example the leading divergence is in general proportional to the area of the entangling surface \cite{Srednicki:1993im}. In conformal field theories the coefficient of the logarithmically divergent piece in the entanglement entropy is also universal, related to the conformal anomaly of the theory. There can be similar universal terms in general field theories. The finite part of the entanglement entropy can also be universal. So it seems that  there may be a way of understanding the structure of entanglement entropy which does not depend on the details of the field theory under consideration. This is analogous to the case of anomalies in field theories. The structure of the anomaly does not depend on the details of the field theory. One can determine the structure by solving, for example, the Wess-Zumino consistency condition \cite{Wess:1971yu}. The coefficients multiplying these anomalous terms certainly depend on the specific field theory. If a similar procedure exists in the case of entanglement entropy, then that will be a useful tool for practical computations. 

In this brief note we would like to point out that the entanglement entropy in diffeomorphism invariant field theories\footnote{By this we mean that the field theory has a conserved stress-energy tensor and so can be consistently coupled to gravity.} have to satisfy quite strong constraints. One can derive these constraints in the same way as Ward identities are derived in field theory. The only difference is that in this case the path integral is over field configurations which satisfy nontrivial boundary condition and as a result the final answer is similar to an "anomalous Ward identity". In particular, it has to satisfy a Wess-Zumino like consistency condition.

There are some similarities between Wilson loop expectation value  and entanglement entropy in field theory. When we compute the entanglement entropy using the replica trick we essentially compute a path integral over a specific set of field configurations which are chosen in such a way that the entangling surface plays a distinguished role. In the case of wilson loops things are simpler because a particular loop in the space-time is picked up by explicit operator insertion. But in both cases the underlying  diffeomorphism invariance of the field theory is broken by the choice of the entangling surface or the loop. By this we mean the following. The partition function or the quantum effective action of the field theory is invariant under the substitution: $g\rightarrow f^{*}g$, where $f$ is the diffeomorphism and $f^{*}g$ is the pulled-back metric. But the entanglement entropy or the Wilson loop expectation value is not invariant under the substitution, because the diffeomorphism also changes the shape of the entangling surface or the loop. So the group of diffeomorphisms will act nontrivially on the entanglement entropy or the Wilson loop expectation value. In particular they have to transform in such a way that the algebra of diffeomorphism is satisfied and this will impose some constraints on the possible forms of the entanglement entropy or Wilson loop expectation value.

\section{Entanglement Entropy}
Let us take the background geometry to be of the form $R^{1}\times M^{n+1}$ where $R^{1}$ is the time direction and $M^{n+1}$ is the spatial section. We divide $M^{n+1}$ into two regions, $A$ and $B$, by introducing a codimension one hypersurface ${\Sigma}$ in $M^{n+1}$. We are interested in computing the entanglement entropy of the region $A$. We shall refer to the points belonging to $A$ as points inside ${\Sigma}$. \\
Now we apply a spatial diffeomorphism, $f$, to the background geometry. The background metric is of the form:
\begin{equation}
{ds}^{2} = g_{\mu\nu}{dx}^{\mu}{dx}^{\nu} = - {dt}^{2} + {\gamma}_{ij}(x) {dx}^{i}{dx}^{j}
\end{equation}
where $t$ is the time coordinate and ${\gamma}_{ij}$ is the time independent positive definite spatial metric on the constant time slices, $M^{n+1}$. Since $f$ is purely spatial it essentially acts only on the spatial metric ${\gamma}_{ij}$ and its action does not depend on the choice of the time slice. In general the diffeomorphism maps the regions $A$ and $B$ to some other regions $f(A)$ and $f(B)$. The boundary ${\Sigma}$ between $A$ and $B$ gets mapped to $f({\Sigma})$ which is again the boundary between $f(A)$ and $f(B)$. We shall assume that the diffeomorphism is continuously connected to the identity and hence can be thought of as generated by some vector field. As a result of this  the diffeomorphism never carries any point of region $A$ or $B$ across the boundary ${\Sigma}$.  In short, reflections across the boundary ${\Sigma}$ are not allowed and orientation is preserved by the diffeomorphism. This together with the fact that the diffeomorphism is time independent has the following interesting consequence. If ${\Phi}$ is a field configuration contributing to the path integral which computes the entanglement entropy of $A$, then $(f^{-1})^{*}{\Phi}$ is another field configuration which contributes to the path integral which computes the entanglement entropy of $f(A)$. In fact this leads to the identity $S_{E}(f(A),g) = S_{E}(A, f^{*}g)$. In this equation $g$ is the background metric and $f^{*}g$ denotes the pull-back of the background metric by the diffeomorphism $f$. $S_{E}(A,g)$ is the entanglement entropy of the region $A$ with the background metric $g$. If ${\xi}$ is the vector field which generates the diffeomorphism then $(f^{*}g)_{\mu\nu} = g_{\mu\nu}+ {\nabla}_{\mu}{\xi}_{\nu} + {\nabla}_{\nu}{\xi}_{\mu}$. In our setting the vector field ${\xi}$ is time independent. In the following section we shall give a heuristic derivation of the above relation.\\\\

We start with the definition of the entanglement entropy of region $A$,
\begin{equation}
S_{A} = -Tr_{H_{A}}{\rho}_{A}ln{\rho}_{A} = - \lim_{n\to 1} \frac{\partial}{\partial n} Tr_{H_{A}}{{\rho}_{A}}^{n}
\end{equation}
In practice the trace over the reduced density matrix ${\rho}_{A}$ cannot be calculated for arbitrary $n$ and so one computes it for integer values of $n$ and then analytically continue it to arbitrary values of $n$. But it should be mentioned that the existence and uniqueness of a proper analytic continuation cannot always be proved. So let us study the quantity $Tr_{H_{A}}{{\rho}_{A}}^{n}$ for integer values of $n$ using path integral.

The matrix element of the reduced density matrix ${\rho}_{A}$ is defined as,
\begin{equation}
{\rho}_{A}({\phi}_{A},{\phi}'_{A}) |_{g} = Tr_{H_{B}} {\rho} |_{g} = \int D_{g}{{\phi}}_{B} \  {<{\phi}_{A},{\phi}_{B}| {\rho} |{\phi}'_{A},{\phi}_{B}>_{g} }
\end{equation}
In the above equation ${\phi}$ denotes all the fields in the theory. We have decomposed the field eigenstate $|{\phi}>$ on the spatial section $M^{n+1}$ as $|{\phi}> = |{\phi}_{A}>\otimes |{\phi}_{B}>$ and $D_{g}{\phi}_{B}$ is the integration measure over the field configurations ${\phi}_{B}$. We have also introduced the background metric $g$ on $M^{n+1}$ and all the integration measures depend on it. Using the above definition we can write,
\begin{align}
Tr_{H_{A}} {{\rho}_{A}}^{2} |_{g} &=  \int D_{g} {{\phi}_{A}}D_{g} {{\phi}'_{A}} \ {\rho}_{A}({\phi}_{A},{\phi}'_{A}){\rho}_{A}({\phi}'_{A},{\phi}_{A})\\
&= \int D_{g} {{\phi}_{A}}D_{g} {{\phi}'_{A}} D_{g} {{\phi}_{B}}D_{g} {{\phi}'_{B}} <{\phi}_{A},{\phi}_{B}| {\rho} |{\phi}'_{A},{\phi}_{B}>_{g} <{\phi}'_{A},{\phi}'_{B}| {\rho} |{\phi}_{A},{\phi}'_{B}>_{g} \nonumber
\end{align}
Now the density matrix of the system is given by, ${\rho} = \frac{e^{-{\beta}H}}{Z}$, where ${\beta}$ is the inverse temperature, $H$ is the Hamiltonian and $Z$ is the canonical partition function of the system at temperature ${\beta}^{-1}$. We can write matrix elements as,
\begin{equation}
<{\phi}_{A},{\phi}_{B}| {\rho} |{\phi}'_{A},{\phi}_{B}>_{g} \ = \ \frac{1}{Z} \int_{{\phi}(0) = ({\phi}'_{A},{\phi}_{B})}^{{\phi}({\beta}) = ({\phi}_{A},{\phi}_{B})}  D_{g}{\phi} \ e^{-S({\phi,g})}
\end{equation}
where $S$ is the Euclidean action of the field theory. Now we make a change of variable in the path integral, ${\phi} \rightarrow {\tilde {\phi}}$ ($= (f^{-1})^{*}{\phi} $),
\begin{align}
<{\phi}_{A},{\phi}_{B}| {\rho} |{\phi}'_{A},{\phi}_{B}> _{g}& \ =\  \frac{1}{Z} \int_{\tilde{{\phi}}(0) = ({\phi}'_{A},{\phi}_{B})}^{\tilde{{\phi}}({\beta}) = ({\phi}_{A},{\phi}_{B})}  D_{g} \tilde{{\phi}} \ e^{-S(\tilde {{\phi}},g)}  \\
&= \frac{1}{Z} \int_{(f^{-1})^{*}{\phi}(0) = ({\phi}'_{A},{\phi}_{B})}^{(f^{-1})^{*}{\phi}({\beta}) = ({\phi}_{A},{\phi}_{B})}  D_{g}(f^{-1})^{*}{\phi} \ e^{-S((f^{-1})^{*}{\phi},g)} \nonumber \\
&= \frac{1}{Z} \int_{{\phi}(0) = (f^{*}{\phi}'_{A},f^{*}{\phi}_{B})}^{{\phi}({\beta}) = (f^{*}{\phi}_{A},f^{*}{\phi}_{B})}  D_{f^{*}{g}}{\phi} \ e^{-S({\phi},f^{*}g)} \nonumber
\end{align}
where $f$ is a diffeomorphism which acts only on the spatial section $M^{n+1}$ and is time independent. In the last line we have used the diffeomorphism invariance of the action and the path integral measure. We can write this as an identity,
\begin{equation}
<{\phi}_{A},{\phi}_{B}| {\rho} |{\phi}'_{A},{\phi}_{B}>_{g} \ = \ <f^{*}{\phi}_{A},f^{*}{\phi}_{B}| {\rho} |f^{*}{\phi}'_{A},f^{*}{\phi}_{B}>_{f^{*}g}
\end{equation}
where we have used the fact that the partition function $Z$ is diffeomorphism invariant. Substituting this identity into eqn-(4) we get,
\begin{equation}
\begin{split}
Tr_{H_{A}} {{\rho}_{A}}^{2} |_{g} &= \int D_{g} {{\phi}_{A}}D_{g} {{\phi}'_{A}} D_{g} {{\phi}_{B}}D_{g} {{\phi}'_{B}} <f^{*}{\phi}_{A},f^{*}{\phi}_{B}| {\rho} |f^{*}{\phi}'_{A},f^{*}{\phi}_{B}>_{f^{*}g}\\
& \quad \times <f^{*}{\phi}'_{A},f^{*}{\phi}'_{B}| {\rho} |f^{*}{\phi}_{A},f^{*}{\phi}'_{B}>_{f^{*}g}
\end{split}
\end{equation}
Now we use the diffeomorphism invariance of the measure and write,
\begin{equation}
\begin{split}
Tr_{H_{A}} {{\rho}_{A}}^{2} |_{g} &= \int D_{f^{*}g} {f^{*}{\phi}_{A}}D_{f^{*}g} {f^{*}{\phi}'_{A}} D_{f^{*}g} {f^{*}{\phi}_{B}}D_{f^{*}g} {f^{*}{\phi}'_{B}} <f^{*}{\phi}_{A},f^{*}{\phi}_{B}| {\rho} |f^{*}{\phi}'_{A},f^{*}{\phi}_{B}>_{f^{*}g} \\
& \quad  \times <f^{*}{\phi}'_{A},f^{*}{\phi}'_{B}| {\rho} |f^{*}{\phi}_{A},f^{*}{\phi}'_{B}>_{f^{*}g} \\
& = Tr_{H_{f^{-1}(A)}} {{\rho}_{f^{-1}(A)}}^{2} |_{f^{*}g}
\end{split}
\end{equation}
This equation follows from the facts that $f^{*}{\phi}_{A}$ and $f^{*}{\phi}_{B}$ are field configuration on the regions $f^{-1}(A)$ and $f^{-1}(B)$ respectively and we are integrating over them with $f$ held fixed. It is clear from the above derivation that this is true for any integer $n$ and so it is a reasonable assumption that this relation still holds after analytic continuation. So differentiating this relation with respect to $n$ we get,
\begin{equation}
S_{E}(A,g) \ = \ S_{E}(f^{-1}(A),f^{*}g)
\end{equation}
or
\begin{equation}
S_{E}(f(A),g) \ = \ S_{E}(A,f^{*}g)
\end{equation}
Although we have derived this relation under certain assumptions, our claim is that this will hold even in those cases where these assumptions are not valid. In other words, \emph{we claim that in any diffeomorphism invariant field theory, the entanglement entropy will satisfy the relation $S_{E}(f(A),g) \ = \ S_{E}(A,f^{*}g)$, where $f$ is any time-independent spatial diffeomorphism which is continuously connected to the identity, i.e, generated by some smooth vector field.} This relation may still be valid for arbitrary diffeomorphism transformations but it certainly does not follow from the above logic. 

It also follows from the above equality that the change in the entanglement entropy when the shape of the region is deformed is captured completely by the stress-tensor of the theory. We showed that changing the shape of the entangling surface is equivalent to changing the background metric keeping the shape of the entangling surface fixed. But the response to a change in the background metric is captured by the stress tensor of the theory.

Now we can expand the entanglement entropy as :
\begin{equation}
S_{E}(f(A),g) = S_{E}(A,f^{*}g) = S_{E}(A,g) + \int{d^{n+1}x\sqrt{g}{\delta}g_{\mu\nu}}\frac{1}{\sqrt g}\frac{{\delta}S_{E}}{{\delta}g_{\mu\nu}} 
\end{equation}
If we use the above formula for the change in the metric we get:
\begin{equation}
S_{E}(f(A),g) = S_{E}(A,f^{*}g) = S_{E}(A,g) + \int{d^{n+1}x \sqrt g {\xi}_{\mu}{\nabla}_{\nu}(-\frac{2}{\sqrt{g}}\frac{{\delta} S_{E}}{{\delta} g_{\mu\nu}})}
\end{equation}
It is easy to show by studying diffeomorphisms for which $f(A) = A$, that the integrand has a delta function support on the entangling surface. It is not necessary for $f$ to keep $A$ point-wise fixed. In other words the diffeomorphism has to be such that the vector field which generates this has to either vanish on the boundary of $A$ or should be tangential to the boundary at every point. This is equivalent to the condition $f(A) = A$. So we can write ,
\begin{equation}
{\nabla}_{\nu}(\frac{-2}{\sqrt g}\frac{{\delta}S_{E} }{{\delta} g_{\mu\nu}}) = F n^{\mu} {{\delta}_{\Sigma}}
\end{equation}
where $n^{\mu}$ is the unit outward normal to the surface ${\Sigma}$ and ${\delta}_{\Sigma}$ is the delta function supported on the surface ${\Sigma}$\footnote{${\delta}_{\Sigma}$ is defined the following integral identity: $\int_{M} \sqrt{g}{f {\delta}_{\Sigma}} = \int_{\Sigma}{\sqrt{h}f_{{\Sigma}}}$ where $M$ is the space-time, ${\Sigma}$ is a codimension one hypersurface, $f$ is a scalar function, $g$ is the background metric ,$h$ is the induced metric on the surface ${\Sigma}$ and ${f_{\Sigma}}$ is the scalar function $f$ evaluated on the surface ${\Sigma}$. }.As a result we can write, 
\begin{equation}
{\delta}_{\xi}S_{E}(A,g) = S_{E}(A,f^{*}g) - S_{E}(A,g) = \int_{\Sigma}{F{\xi}^{{\mu}}n_{{\mu}}}
\end{equation}
The right hand side of eqn-(14) could contain derivatives of the delta function of the form $n^{\mu}{{\nabla}_{\mu}}{\delta}_{\Sigma}$, but they can be eliminated by the following argument. If it contains derivatives of the delta function then the change in the entanglement entropy under a deformation of the entangling surface will contain a term proportional to $n^{\mu}{\nabla}_{\mu}{\xi}_{\nu}$ evaluated on the surface ${\Sigma}$. Now we can choose ${\xi}$ such that it is vanishing along the surface ${\Sigma}$ but has a nonzero normal derivative along ${\Sigma}$. In that case $f(A)= A$ and using eqn-(13) we get an equation of the form,
\begin{equation}
\int{d^{n}x\sqrt{h}F'n^{\mu}n^{\nu}{\nabla}_{\nu}{\xi}_{\mu}} = 0
\end{equation} 
This has to be satisfied for all ${\xi}$ which vanish on the surface ${\Sigma}$. But we can choose the normal derivative of ${\xi}$ on the surface arbitrarily even if we keep ${\xi}$ fixed on the surface. So the only way this can vanish is if $F' = 0$. So we do not need any derivative of delta function on the right hand side.The fact that the R.H.S is proportional to the normal vector,follows from the invariance of the entanglement entropy under the reparametrization of the surface ${\Sigma}$. $F$ is some scalar function constructed out of the metric , the normal vector field and their derivatives. \\\\
If we are computing the entanglement entropy in a pure state then we must have $S_{E} (A,g) = S_{E}(B,g)$.  This requires that the function $F$ should be odd under the change $n^{\mu} \rightarrow -n^{\mu}$ where $n^{\mu}$ is unit normal to the entangling surface. We follow the convention that the unit normal points in the direction away from the region for which we are computing the entanglement entropy. \\\\
We can see that even eqn-(15) is strong enough to rule out many possible terms in the entanglement entropy. For example consider a term in the expression for the entanglement entropy which can be written as an integral over the entangling surface. If we evaluate the left hand side of eqn-(15) on this term then generically this will produce terms containing normal derivatives of the vector field on the surface ${\Sigma}$ and this cannot be integrated out because we have a surface integral. If this is the case then that term alone cannot appear in the entanglement entropy. So we can either drop that term or if possible add some other terms to the entanglement entropy which will cancel the derivatives of the delta function.
It is easy to show that the Area of the surface ${\Sigma}$ is a consistent solution of this equation. For the surface area contribution, $F = K_{\Sigma}$, where $K_{\Sigma}$ is the trace of the extrinsic curvature of $\Sigma$. We would like to emphasize that we have not proved the area law. We have shown that in any diffeomorphism invariant field theory area-law is one of the consistent solutions.\footnote{ This equation is very similar to the loop equation for wilson loops. One can show that the area law in the confining phase is a consistent solution of the loop equation, but it is difficult to show that the string tension vanishes if there is no confinement, just from the loop equation.  }\\\\
In the next section we shall show that the Wilson loop expectation value satisfies an identical type of equation. We choose this as an example because this is somewhat cleaner and has many things similar to the entanglement entropy. 
\section{Wilson Loop}
In this section we shall show that the Wilson loop expectation value in gauge theory satisfies a similar equation. It will make some of our previous arguments clearer. For the case of Wilson loops we replace our earlier product geometry with an arbitrary Riemannian manifold. For the sake of simplicity we shall also restrict ourselves to Wilson loops in Abelian gauge theories. But everything goes through essentially unchanged for non-Abelian loops. The expectation value of the Wilson loop is given by, 
\begin{equation}
W(C,g) = \int \!{D_{g}A} \ exp(i\int_{C}{A})\ e^{-S(A,g)}
\end{equation} 
were $C$ is a closed contour and $S(A,g)$ is the Euclidean action of the theory. We have introduced a background metric $g$. We further assume that the action $S(A,g)$ and the path integral measure $D_{g}A$ are both diffeomorphism invariant. Now we make a change of variable in the path integral,
\begin{align}
W(C,g)& = \int \!{D_{g}A'} \ exp(i\int_{C}{A'})\ e^{-S(A',g)}\\
&=  \int \!{D_{g}f^{*}A} \ exp(i\int_{C}{f^{*}A})\ e^{-S(f^{*}A,g)}  \nonumber \\
& =  \int \!{D_{(f^{-1})^{*}g}A} \ exp(i\int_{f(C)}{A})\ e^{-S(A,(f^{-1})^{*}g)}  \nonumber \\
& = W(f(C),(f^{-1})^{*}g) \nonumber
\end{align}
where $f$ is a diffeomorphism.  In the first line we have changed the variable of of integration from $A$ to $A'$, in the second line $A'$ is identified with $f^{*}A$ and in the third line we have used the diffeomorphism invariance of the action and the path-integral measure. This equation can also be written as,
\begin{equation}
W(f(C),g) = W(C,f^{*}g)
\end{equation}
This equation tells us that changing the shape of the loop is equivalent to changing the background metric keeping the loop fixed. The change in the background metric is equivalent to stress tensor insertion. In fact it is easy to see that the change is proportional to the correlation function of the Wilson loop operator and operators of the form ${\xi}_{\mu}{\nabla}_{\nu}T^{\nu\mu}$, where $T^{\mu\nu}$ is the stress tensor and ${\xi}$ is the generator of the diffeomorphism. The stress tensor is covariantly conserved and so the only contribution comes from the contact term which is localized on the Wilson loop. The same argument goes through for non-Abelian Wilson loops. For simplicity let us consider only non self-intersecting or simple loops. Since we are only considering diffeomorphisms, simple loops get mapped to simple loops and the orientation is preserved by the diffeomorphism according to our assumption. \\
We now expand the Wilson loop expectation value as,
\begin{equation}
W(f(C),g) = W(C,f^{*}g) = W(C,g) + \int{\mathrm d^{n}x}\sqrt{g}\ {\xi_{b}}{\nabla}_{a}({\frac{-2}{\sqrt{g}} \frac{{\delta}W}{{\delta}g_{ab}})}
\end{equation}
where ${\xi}$ is the vector field generating the diffeomorphism $f$. So as long as the diffeomorphism does not move the loop, i.e, $f(C)=C$ ,we get
\begin{equation}
\int{\mathrm d^{n}x}\sqrt{g}\ {\xi_{b}}{\nabla}_{a}({\frac{-2}{\sqrt{g}} \frac{{\delta}W}{{\delta}g_{ab}})} = 0
\end{equation}
then by standard arguments, we get 
\begin{equation}
{\nabla}_{a}({\frac{-2}{\sqrt {g}}} \frac{{\delta}W}{{\delta}g_{ab}}) = F^{(i)}n^{b}_{(i)}{\delta}_{C}
\end{equation} 
where the $n_{(i)}$-s are $(n-1)$ orthonormal vectors all of which are orthogonal to the loop $C$. The right hand side is proportional to the vectors $n_{i}$ because the expectation value is invariant under the reparametrizations of the loop. $F^{(i)}$-s are $(n-1)$ scalar functions built out of the metric, the normal vectors and their derivatives. $F^{i}$ must transform in the vector representation of the local $SO(n-1)$ group which acts on the set of orthonormal vectors, $n_{i}$. This follows from the redundancy in our choice of the normal vectors which are defined only up to local $SO(n-1)$ transformation. The change of $W(C,g)$ under a deformation of the loop $C$ can be written as,
\begin{equation}
W(f(C),g) = W(C,f^{*}g) = W(C,g) + \int_{C}{{\xi}_{b}F^{i}n^{b}_{i}}
\end{equation}
\subsection{W Vs. lnW}
So far we have not talked about the nature of the functions $F^{i}$. In general they will not be local, i.e, a scalar function built out of the metric, the normal vectors $n_{i}$ and their derivatives evaluated at a given point of the space. In particular the variation of the Wilson loop will produce functions which will not have this property in general. But $lnW$ is expected to be much better behaved in this respect. Our previous analysis remains unchanged if we replace $W$ with $lnW$ and so we can write,
\begin{equation}
lnW(f(C),g) =ln W(C,f^{*}g) = lnW(C,g) + \int_{C}{{\xi}_{b}F^{i}n^{b}_{i}}
\end{equation}
where the functions $F^{i}$ are expected to contain local terms. It can also contain non local terms whose existence cannot be ruled out by this procedure. 
\section{Wess-Zumino Like Integrability Condition}
In the previous sections we have studied entanglement entropy  and Wilson loop expectation values in arbitrary field theories. We have shown that they both behave similarly when the shape of the entangling surface or the loop is deformed. In both cases we have written down the variation in terms of some unknown functions. But is there any way to determine them? Entanglement entropy (Wilson loop expectation value) or more precisely the function $F (F^{i})$ has to satisfy an integrability condition. Eqn-(2) gives the change of entanglement entropy under a diffeomorphism. But \emph{given two regions there exist in general more than one (infinitely many) diffeomorphisms connecting them and all of them should give the same change in entanglement entropy}. This is equivalent to the condition that  the algebra of diffeomorphism has to be obeyed. It leads to the following condition,
\begin{equation}
[{\delta}_{{\xi}_{1}},{\delta}_{{\xi}_{2}}] S_{E} = {\delta}_{{[{\xi}_{2},{\xi}_{1}]}} S_{E}
\end{equation}
The vector fields in the commutator appear in the reverse order because we are considering active diffeomorphism.
One can write eqn(5) in terms of the function $F$ as
\begin{equation}
{\delta}_{{\xi}_{1}}Q({{\xi}_{2}},F) - {\delta}_{{\xi}_{2}}Q({\xi}_{1},F) = Q({[{\xi}_{2},{{\xi}_{1}}]},F)
\end{equation}
where $Q$ is the integral
\begin{equation}
Q({\xi},F) = \int_{\Sigma}{\mathrm{{d^{n}x}} \sqrt h\ n_{\mu}{\xi}^{\mu}F}
\end{equation}
Here $h$ is the induced metric on the entangling surface ${\Sigma}$. This is our main result. We have checked that $F=K_{\Sigma}$ which comes from the area term in the entanglement entropy satisfies this integrability condition. It is easy to check that any arbitrary function $F$ will not satisfy this integrability condition. So this condition is nontrivial. Similarly we can write down the same consistency condition for the Wilson loop.
\subsection{Comments on The Solution of The Integrability Condition}
We hope that the integrability condition we have stated above will be solvable in the same way as the Wess-Zumino consistency condition. But the difference with the standard anomaly answer is clearly visible here. In our case the "anomaly" is localized on a finite codimension (entangling) surface. Also in the case of anomaly in field theory, the gauge variation of the quantum effective action is always local and finite. But in this case it is not clear if the function $F(F^{i})$ is always local. In general it will contain both local and non-local terms. Also the entanglement entropy and the Wilson loop expectation values are both ultraviolet divergent quantities. If we assume that the field theory has been regularized in such a way that diffeomorphism invariance is respected, then this should not cause any problem. The interesting point is that one can determine the structure of the local terms in $F(F^{i})$ by solving the consistency condition. There will be two kinds of solutions. The trivial solutions will be those which can be obtained by applying a diffeomorphism to the integral of a \emph{local} function. The integral can be a surface integral on ${\Sigma}$ or a volume integral defined on the region bounded by ${\Sigma}$. But the interesting solutions are those which cannot be obtained in this way and presumably there is only a finite number of solutions with this property\footnote{There are only a finite number of them because in general the solutions of this problem are the cohomology classes of a BRST like operator.}. In other words,\emph{ the nontrivial solutions of the integrability condition define the nontrivial local terms which can appear in the variation of the entanglement entropy}. In many cases the variation of the entanglement entropy is more well defined than the entanglement entropy itself because the ultraviolet cutoff may cancel in the variation. The solution of the integrability condition we have stated above will give the nontrivial local terms which can appear in the variation. It is important to note that in order to solve this integrability condition one  does not need to know about the field theory, except that the specific numerical coefficients multiplying the purely geometric expressions will depend on the theory. It may happen that some of the local contributions in $F$ are actually protected. This may provide us with some examples of protected quantities in non-supersymmetric field theory.  The other point is that  this method does not require introduction of a background geometry with conical defect and so the calculation is under control at every stage.\\\\

{\bf Acknowledgement:} I am grateful to Shamit Kachru, Shunji Matsuura, Ashoke Sen, Steve Shenker and Leonard Susskind for very useful discussion on these subjects. I would also like to thank Ashoke Sen and Steve Shenker for helpful comments on the draft.

\end{document}